\documentclass[USenglish,
10pt,twocolumn]{article}
\usepackage[latin1]{inputenc}
\usepackage
{graphicx}
\usepackage{epstopdf}
\usepackage{enumerate}
\usepackage{amsmath}
\usepackage{amssymb}

\usepackage{vmargin}
\setpapersize{A4}
\setmarginsrb{2.5cm}{0.5cm}{2.5cm}{3.5cm}{1.5cm}{1cm}{1.5cm}{1.5cm}

\newcommand{\C}{\mathcal{C}}

\newcommand{\Cs}{\mathbf{C}} 
\newcommand{\SC}{\mathcal{A}}
\newcommand{\Ss}{\mathbf{A}}

\newcommand{\N}{\mathbb{N}}

\newcommand{\Link}{\text{L}}

\newcommand{\T}[1]{\texttt{#1}}
\newcommand{\ED}[1]{E_{#1}}

\newcommand{\squote}[1]{\emph{``#1''}}

\begin{document}

\title{Measuring Generalized Preferential Attachment in Dynamic Social Networks
\author{Camille Roth\footnote{CREA (Center for Research in Applied
    Epistemology), CNRS/Ecole Polytechnique, 1 rue Descartes, 75005 Paris, France.
     E-mail: \T{roth@shs.polytechnique.fr} }  } }

\maketitle 
\begin{abstract}{\small The mechanism of preferential attachment underpins most recent social network formation models. Yet few authors attempt to check or quantify assumptions on this mechanism. We call \emph{generalized preferential attachment} any kind of preference to interact with other agents with respect to any node property. We then introduce tools for measuring empirically and characterizing comprehensively such phenomena, and apply these tools to a socio-semantic network of scientific collaborations, investigating in particular homophilic behavior. This opens the way to a whole class of realistic and credible social network morphogenesis models.
\vspace{0.5cm}

}

\vspace{0.7cm}
{\noindent \scriptsize ACM: G.2.2; MSC: 68R10; PACS: 89.65.-s, 87.23.Ge, 89.75.-k}

\end{abstract}

\vspace{0.5cm}



\section*{Introduction}


A recent challenge in structural research in social science consists in modeling social network formation. Social networks are usually interaction networks --- nodes are agents and links between nodes represent interactions between agents --- and in this respect, modeling them involves disciplines linked both to graph theory (computer science and statistical physics), mathematical sociology and economics \cite{bara:stat,cohe:emer,BrianSkyrms08012000}. Most of the interest in this topic stems from the empirical observation that real social networks strongly differ from uniform random graphs as regards several statistical parameters; and foremost with respect to node connectivity distribution, or \emph{degree} distribution. Indeed, in random graphs \emph{a la} Erdos-Renyi \cite{erdo:rand} links between agents are present with a constant probability $p$ and degree distributions follow a Poisson law whereas empirical social networks exhibit power-law, or \emph{scale-free}, degree distributions 
\cite{bara:stat}. 
This phenomenon suggested that link formation does not occur randomly but instead depends on node and network properties --- 
that is, agents do not interact at random but instead according to heterogeneous preferences for other nodes. 

Hence, early social network models endeavored to describe non-uniform interaction and growth mechanisms yielding the famous ``scale-free'' degree distribution \cite{bara:emer}. Subsequently, much work has been focused on determining processes explaining and rebuilding more complex network structures consistent with those observed in the real world --- a consistency validated through a rich set of statistical parameters measured on empirical networks, not limited to degree distribution but including as well clustering coefficient, average distance, assortativity, etc. \cite{bogu:mode,cald:scal,newm:stru}. 

However, even when cognitively, sociologically or anthropologically credible, most of  the hypotheses driving these models are mathematical abstractions that do not enjoy any experimental measurement or justification.
In this paper, we call \emph{preferential attachment} (PA) {any kind} of non-uniform interaction behavior and introduce tools for empirically measuring PA with respect to any node property. We eventually apply these measures to an empirical case of socio-semantic network. In particular, we will criticize degree-related PA, and estimate homophily.

\section{A brief survey of social network models}

Barabasi \& Albert \cite{bara:emer} pioneered the use of preferential linking in social network formation models  to successfully rebuild a particular statistical parameter, the scale-free  degree distribution. In their model, new nodes arrive at a constant rate and attach to already-existing nodes with a likeliness linearly proportional to their degree.  This model has been widely spread and reused, and consequently made the term ``preferential attachment'' often understood as  degree-related only preferential attachment. 
Since then, many authors introduced diverse modes of preferential link creation depending on either various node properties (hidden variables and ``types'' \cite{bogu:clas,sode:gene}, fitness~\cite{cald:scal}, centrality, euclidian distance \cite{fabr:heur}, common friends \cite{jin:stru}, bipartite structure \cite{guil:bipa}, \hbox{etc.}) or on various linking mechanisms (competitive trade-off and optimization heuristics~\cite{colizza:network,fabr:heur}, two-steps node choice~\cite{stef:pref}, to cite a few). 

However and even in recent papers, hypotheses on PA are often arbitrary and at best supported by qualitative intuitions. Existing \emph{quantitative} estimations of PA and consequent validations of modeling assumptions are extremely rare. In this respect, most studies are either (i) related to the classical degree-related PA~\cite{bara:evol,eise:pref,jeong:measurin,redn:cita}, sometimes extended to a selected network property, like common acquaintances~\cite{newm:clus}; or (ii) reducing PA to a single parameter: for instance using econometric approaches~\cite{powe:grow} or Markovian models \cite{snij:stat}. 
While of great interest in approaching the underlying behaviorial reality of social networks, these works may not be able to provide a sufficient empirical basis and support for designing trustworthy PA mechanisms, and accordingly for proposing credible social network morphogenesis models.
Yet in this view we argue that the three following points are key:
\begin{enumerate}
\item Node degree does not make it all --- and even the popular degree-related PA (a linear ``rich-get-richer'' heuristics) seems to be inaccurate for some types of real networks \cite{bara:evol}, and possibly based on flawed behavioral fundations, as we will suggest below in Sec. \ref{sec:barabash}.
\item Strict social network topology and derived properties may not be sufficient to account for complex social phenomena --- as several above-cited models suggest, introducing ``external'' properties (such as \hbox{e.g.} node types) may influence interaction; explaining for instance homophily-related PA \cite{mcph:homo} requires at least to \emph{qualify} nodes.
\item Single parameters cannot express the rich heterogeneity of interaction behavior --- for instance, when assigning a unique 
 parameter 
  to preferential interaction with close nodes, one misses the fact that such interaction could be significantly more frequent for very close nodes than for loosely close nodes, or be quadratic with respect to the distance, etc. 
\end{enumerate}

To summarize, it is thus crucial to conceive PA in such a way that (i) it is a flexible and general mechanism, depending on relevant parameters based on both topological and non-topological properties; and (ii) it is an empirically valid function describing the whole scope of possible interactions.

\section{Measuring preferential attachment}
PA is the likeliness for a link to appear between two nodes with respect to node properties. In order to measure it, we first have to distinguish between (i) single node properties, or \emph{monadic} properties  (such as degree, age, etc.) and (ii) node dyad properties, or \emph{dyadic} properties (social distance, dissimilarity, etc.).
When dealing with monadic properties indeed, we seek to know the propension of some kinds of nodes to be involved in an interaction. On the contrary when dealing with dyads, we seek to know the propension for an interaction to occur preferentially with some kinds of couples.\footnote{Note that a couple of monadic properties can be considered dyadic; for instance, a couple of nodes of degrees $k_{1}$ and $k_{2}$ considered as a dyad $(k_{1},k_{2})$. This makes the former case a refinement, not always possible, of the latter case.}

\subsection{Monadic PA}
Suppose we want to measure the influence on PA of a given monadic property $m$ taking values in $\mathcal{M}=\{m_{1},..., m_{n}\}$.\footnote{This topic will be described more extensively in a forthcoming general paper presenting basic parameters for dynamic network analysis \cite{lata:bas2}.} We assume this influence can be described by a function $f$ of $m$, independent of the distribution of agents of kind $m$. Denoting by ``$\Link$'' the event ``attachment of a new link'', $f(m)$ is simply the conditional probability $P(\Link|m)$ that an agent of kind $m$ is involved into an interaction. 

Thus, it is $f(m)$ times more probable that an agent of kind $m$ receives a link. We call $f$ the \emph{interaction propension} with respect to $m$.
For instance, the classical degree-based PA used in Barabasi-Albert and subsequent models --- links attach proportionally to node degrees  \cite{bara:emer,bara:evol,cata:asso} --- is an assumption on $f$ equivalent to $f(k)\propto k$. 

$P(m)$ typically denotes the distribution of nodes of type $m$.
The probability $P(m|\Link)$ for a new link extremity to be attached to an agent of kind $m$ is therefore proportional to $f(m)P(m)$, or $P(\Link|m)P(m)$. Applying the Bayes formula yields indeed:\footnote{For consistency purposes, we also assume $f$ strictly positive: $\forall m\in\mathcal{M}, f(m)>0$.}
\begin{equation}P(m|\Link)=\frac{f(m)P(m)}{P(\Link)}\end{equation}
with $P(\Link)=\displaystyle\sum_{m'\in\mathcal{M}}f(m')P(m')$.

Empirically, during a given period of time $\nu$ new interactions occur and  $2\nu$ new link extremities appear. Note that a repeated interaction between two already-linked nodes is not considered a new link, for it incurs acquaintance bias. The expectancy of new link extremities attached to nodes of property $m$ along a period is thus $\nu(m)=P(m|\Link)\cdot2\nu$. 
As $\displaystyle\frac{2\nu}{P(\Link)}$ is a constant of $m$ and  the network is considered  static during the time period, 
  we may estimate $f$ through $\hat{f}$ such that: 
\begin{equation}
\left\lbrace
\begin{array}{lr}
\label{eq:f}
\hat{f}(m)=\displaystyle\frac{\nu(m)}{P(m)}&\text{ if }P(m)>0\\
\hat{f}(m)=0&\text{ if }P(m)=0
\end{array}
\right.
\end{equation}

Thus $\mathbf{1}_{P}(m)f(m)\propto\hat{f}(m)$, where $\mathbf{1}_{P}(m)=1$ when  $P(m)>0$, $0$ otherwise. 

\subsection{Dyadic PA}
Adopting a dyadic viewpoint is required whenever a property has no meaning for a single node, which is mostly the case for properties such as proximity, similarity --- or distances in general. We therefore intend to measure interaction propension for a dyad of agents which fulfills a given property $d$ taking values in $\mathcal{D}=\{d_{1},d_{2},...,d_{n}\}$. Similarly, we assume the existence of an essential dyadic interaction behavior embedded into $g$, a strictly positive function of $d$; correspondingly the conditional probability $P(\Link|d)$. Again, interaction of a dyad satisfying property $d$ is $g(d)$ times more probable. In this respect, the probability for a link to appear between two such agents is:
\begin{equation}P(d|\Link
)=\displaystyle\frac{g(d)P(d)}{P(\Link)}\end{equation}
with $P(\Link)=\displaystyle\sum_{d'\in\mathcal{D}}g(d')P(d')$.

Here, the expectancy of new links between  dyads of kind $d$ is $\nu(d)=P(d|\Link)\nu$. Since 
$\displaystyle\frac{\nu}{P(\Link)}$ is a constant of $d$ we may estimate $g$ with $\hat{g}$:
\begin{equation}
\left\lbrace
\begin{array}{lr}
\label{eq:g}
\hat{g}(d)=\displaystyle\frac{\nu(d)}{P(d)}&\text{ if }P(d)>0\\
\hat{g}(d)=0&\text{ if }P(d)=0
\end{array}
\right.
\end{equation}
Likewise, we have $\mathbf{1}_{P}(d)g(d)\propto\hat{g}(d)$.

\section{Interpreting int\-er\-ac\-tion propensions}
\subsection{Shaping hypotheses}\label{sec:correlation}
The PA behavior embedded in $\hat{f}$ (or $\hat{g}$) for a given monadic (or dyadic) property can be reintroduced as such in modeling assumptions, either (i) by reusing the exact empirically calculated function, or (ii) by stylizing the trend of $\hat{f}$ (or $\hat{g}$) and approximating $f$ (or $g$) by more regular functions, thus making possible analytic solutions. 

Still, an acute precision when carrying this step is often critical, for a slight modification in the hypotheses (e.g. non-linearity instead of linearity) makes some models unsolvable or strongly shakes up their conclusions.
For this reason, when considering a property for which there is an underlying natural order, it may also be useful to examine the cumulative propension $\hat{F}(m_{i})=\displaystyle
\sum_{m'=m_{1}}^{m_{i}}\hat{f}(m')$ as an estimation of the integral of $f$, especially when the data are noisy 
 (the same goes with $\hat{G}$ and $\hat{g}$).

\subsection{Correlations between properties} Besides, if modellers want to consider PA with respect to a collection of properties, they have to make sure that the properties are uncorrelated or that they take into account the correlation between properties: evidence suggests indeed that for instance node degrees depend on age. Often models assume properties to be uncorrelated which, when it is not the case, would amount to count twice a similar effect.\footnote{Like for instance in \cite{jin:stru} where effects related to degree and common acquaintances are combined in an independent way.} 

If two distinct properties $p$ and $p'$ are independent, the distribution of nodes of kind $p$ in the subset of nodes of kind $p'$ does not depend on $p'$, i.e. the quantity $\displaystyle \frac{P(p|p')}{P(p)}$ must theoretically be equal to 1, $\forall p, \forall p'$. Empirically, it is possible to estimate it through:\footnote{For computing the correlation between a monadic and a dyadic property, it is easy to interpret $P(p|d)$ as the distribution of $p$-nodes being part of a dyad $d$.}
\begin{equation}
\left\lbrace
\begin{array}{lr}
\label{eq:c}
\widehat{c_{p'}}(p)=\displaystyle\frac{P(p|p')}{P(p)}&\text{ if }P(p)>0\\
\widehat{c_{p'}}(p)=0&\text{ if }P(p)=0
\end{array}
\right.
\end{equation}
in the same manner as previously.

\subsection{Essential behavior} As such, calculated propensions do not depend on the distribution of nodes of a given type at a given time. In other words, if for example physicists prefer to interact twice more with physicists than with sociologists but there are three times more sociologists around, physicists may well be apparently interacting more with sociologists. Nevertheless, $\hat{f}$ remains free of such biases and yields the ``baseline'' preferential interaction behavior of physicists. 

However, $\hat{f}$ could still depend on global network properties, \hbox{e.g.} its size, or its average shortest path length. Validating the assumption that $\hat{f}$ is independent of \emph{any} global property of the network (notably its topological structure) --- \hbox{i.e.}, that it is an  \emph{entirely essential} property of nodes of kind $p$ ---  would require to compare different values of $\hat{f}$ for various periods and network configurations. Put differently, this entails checking whether the shape of $\hat{f}$ itself is a function of global network parameters.

\subsection{Activity}\label{sec:activity} Additionally, $\hat{f}$ represents equivalently an attractivity or an activity: if interactions occur preferentially with some kinds of agents, it could as well mean that these agents are more attractive or that they are more active. If more attractive, the agent will be interacting more, thus being apparently more active. 
To distinguish between the two effects, it is sometimes possible  to measure independently agent activity, notably when interactions occur during \emph{events}, or when interaction initiatives are traceable (\hbox{e.g.} in a directed network). 

In such cases, the distinction is far from neutral for modeling. Indeed, when considering evolution mechanisms focused not on agents creating links, but instead on events gathering agents (like in \cite{rama:self}), modellers have to be careful when  integrating back into models the observed PA as a behavioral hypothesis. Some categories of agents might in fact be more active and accordingly involved in more events, instead of enjoying more attractivity.

\section{An application to socio-semantic networks}

\subsection{Definitions}
We now apply the above tools to a socio-semantic network, that is, a social network where agents are also linked to semantic items. We examine therein two particular kinds of  PA: (i) PA related to a  monadic property: the node degree; and (ii) PA  linked to a dyadic property: homophily, \hbox{i.e.} the propension of individuals to interact more with similar agents. 
\begin{figure}
\begin{center}\includegraphics[width=5cm]{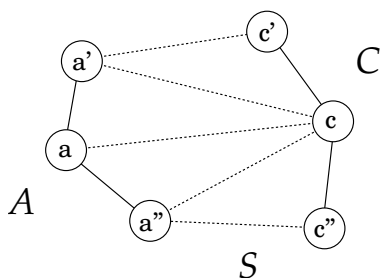}
\caption{Sample socio-semantic network (3 agents $a$, $a'$, $a''$ and 3 concepts $c$, $c'$, $c''$).}
\label{fig:reseau-ex}\end{center}
\end{figure}
\paragraph{Networks.} The social network $\SC$ is the network of agents, where links correspond to interactions: $\SC=(\Ss,\ED{\Ss})$, with $\Ss$ denoting the agent set and $\ED{\Ss}$ the (undirected) set of links between agents.  Interactions occur through events, and each event is associated with a semantic content, made of semantic markers (\hbox{e.g.} keywords), or \emph{concepts}. Dually, the concepts form  a semantic network, where concepts are linked to each other if they jointly appear in an event. Identically to $\SC$, we have $\C=(\Cs,\ED{\Cs})$.  
Finally, agents are linked to concepts associated with events they are involved in. We hence deal with a third  network, $\mathcal{S}=(\Ss,\Cs,\ED{\Ss\Cs})$, and three kinds of links: (i) between pairs of agents, (ii) between pairs of concepts, \hbox{and (iii)} between concepts and agents. Since we  measure agent behavior through network dynamics, we also consider the temporal series of networks $\SC(t)$, $\C(t)$ and $\mathcal{S}(t)$, with $t\in \N$, which altogether make a dynamic socio-semantic network (see Fig.~\ref{fig:reseau-ex}).

\paragraph{Empirical protocol.}
Empirical data come from  the bibliographical database {\sf Medline} which contains dated abstracts of published articles of biology and/or me\-di\-cine. We focused on a portion concerning a well-defined community of embryologists working on the \emph{zebrafish}, during the period 1997-2004. Translated in the above framework,  articles are events, their authors are the agents, and semantic markers are made of expert-selected abstract words.

In order to have a non-empty and statistically significant network for computing propensions, we first build the network on an initialization period of 7 years (from 1997 to end-2003), then carry the calculation on new links appearing during the last year. The dataset contains around $10,000$ authors, $5,000$ articles and $70$ concepts.

\subsection{Linear degree-related PA}\label{sec:barabash} 

We use Eq.~\ref{eq:f} and consider  the node degree $k$ as property $m$ (thus $\mathcal{M}=\N$): in this manner, we intend to compute the real slope $\hat{f}(k)$ of the degree-related PA and compare it with the assumption ``$f(k)\propto k$''. This hypothesis classically relates to the preferential linking of new nodes to old nodes. To ease the comparison, we considered the subset of interactions between a new and an old node.

\begin{figure}\begin{center}
\includegraphics[width=6.2cm]{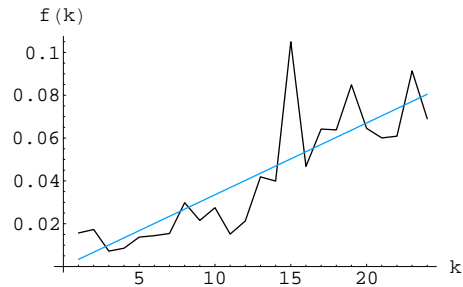}
\end{center}
\caption{\label{fig:kpref}
Degree-related interaction propension $\hat{f}$, computed on a one-year period, for $k<25$. The solid line represents the best linear fit.
}
\end{figure}

\begin{figure}\begin{center}
\includegraphics[width=6.2cm,trim=-2 0 0 0]{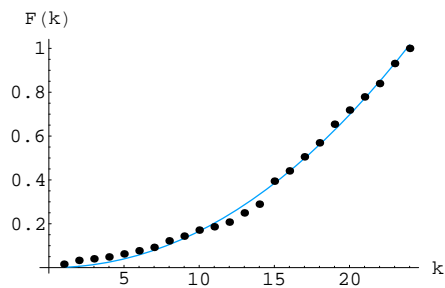}
\end{center}
\caption{\label{fig:cumul} Cumulated propension $\hat{F}$. Dots represent empirical values,  solid color lines are the best non-linear fit for $\hat{F} \sim k^{2.07}$.}
\end{figure}

Empirical results are shown on Fig. \ref{fig:kpref}. Seemingly, the best linear fit corroborates the data and tends to confirm that $f(k)\propto k$. The best non-linear fit however deviates from this hypothesis, suggesting that $f(k)\propto k^{1.09}$. 
As suggested above, knowing precisely the exponent may be critical here. Since there is a natural order on $k$, we plotted the cumulated propension $\hat{F(k)}=\sum_{k'=1}^{k}\hat{f}(k)$ on Fig.~\ref{fig:cumul}. In this case, the best non-linear fit for $\hat{F}$ is $\hat{F}(k)\propto k^{2.07}$, confirming the slight deviation from a strictly linear preference which would yield $k^2$. 

\paragraph{Rich-work-harder.} This precise result is not new and agrees with existing studies of the degree-related PA (e.g. \cite{jeong:measurin,newm:clus}). Nevertheless, we wish to stress a more fundamental point in computing this kind of PA. Indeed, \hbox{considerations} on agent activity lead us to question the usual underpinnings and  justifications of PA  related to a monadic property. 
Concerning degree-related PA, it is the \emph{``rich-get-richer''} me\-ta\-phor describing rich, or well-connected agents as more attractive than poorly connected agents, thus receiving more connections and becoming even more connected.\footnote{\squote{(...) the probability that a new actor will be cast with an established one is much higher than that the new actor will be cast with other less-known actors} \cite{bara:emer}.} 

\begin{figure}\begin{center}
\includegraphics[width=6.2cm]{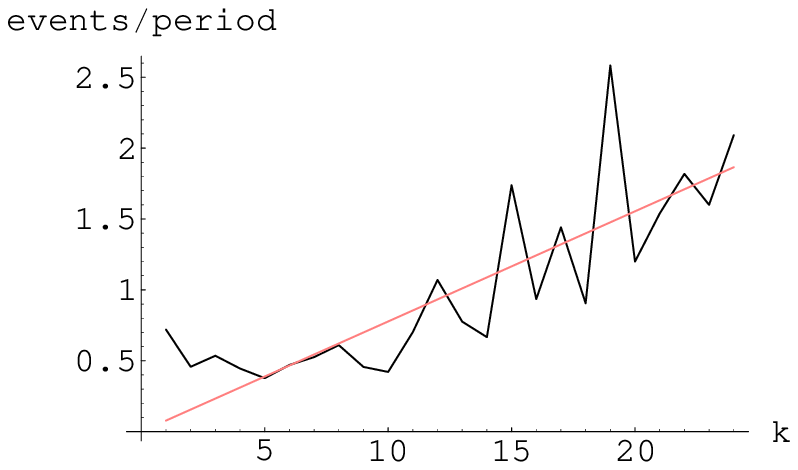}
\vspace{0.3cm}

\includegraphics[width=5.70cm]{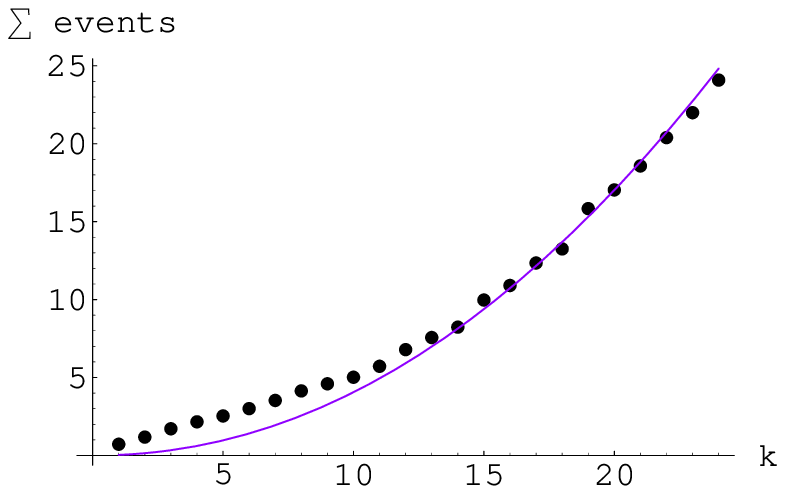}
\end{center}
\caption{\label{fig:activity} \emph{Top:} Activity $a(k)$ during the same period, in terms of articles per period (events per period) with respect to agent degree. Best non-linear fit is $k^{1.02}$. \emph{Below:}  Cumulated activity $A(k)=\sum_{k'=1}^k{a(k)}$.
}
\end{figure}

When considering the activity of agents with respect to $k$, that is, the number of events in which they participate (here, the number of articles they co-author), ``rich'' agents are proportionally more active than ``poor'' agents (see Fig.~\ref{fig:activity}), and thus obviously encounter more interactions. It might thus well simply be that richer agents work harder, not are more attractive; the underlying behavior linked to preferential interaction being simply ``proportional activity''.\footnote{Indeed, when considering $k$ as a proxy for agent activity (i.e., a behavioral feature), if the number of coauthors does not depend on $k$ (which is actually roughly the case in this data) then observing a linear degree-related PA is not surprising.}

While formally equivalent from the viewpoint of PA measurement, the \emph{``rich-get-richer''} and \emph{``rich-work-harder''} me\-ta\-phors are not behaviorally equivalent. One could choose to be blind to this phenomenon and keep an interaction propension proportional to node degree. On the other hand, one could also prefer to consider higher-degree nodes as  more active,  assuming instead that the number of links per event is degree-independent and that agents do neither \emph{prefer}, nor \emph{decide} to interact with famous, highly connected nodes; a hypothesis suggested by the present empirical results. If both of these conceptions are consistent with the observed quasi-proportional PA, they bear different implications for modeling, as underlined in Sec.~\ref{sec:activity}. 

More generally, such feature supports the idea that events, not links, are the right level of modeling for social networks ---  with events reducing in some cases to a dyadic interaction. 

\subsection{Homophilic PA}
Homophily translates the fact that agents prefer to interact with other resembling agents.  Here, we assess the extent to which agents are ``homophilic''  by introducing an inter-agent semantic distance. By semantic distance we mean a function of a dyad of nodes that enjoys the following properties: (i) decreasing with the number of shared concepts between the two nodes, (ii) increasing with the number of distinct concepts, (iii) equal to 1 when agents have no concept in common, and to 0 when they are linked to identical concepts. 

Given $(a, a')\in \Ss^2$ and denoting by $a^\wedge$ the set of concepts $a$ is linked to, we introduce a semantic distance $\delta(a,a')\in[0;1]$ satistying the previous properties:\footnote{Written in a more explicit manner, with $a^\wedge =    \{c_1, ..., c_n, c_{n+1}, ..., c_{n+p} \} $ and $a'^\wedge=    \{c_1,...,c_n,c'_{n+1},...,c'_{n+q}\}$, we have $
\delta(a,a')=\frac{p+q}{p+q+n}$; $n$ and $p$, $q$ representing   respectively the number of elements $a^\wedge$ and $a'^\wedge$ have in common and have in proper. We also verify that if $n=0$ (disjoint sets), $\delta(a,a')=1$; if $n\not =0$, $p=q=0$ (same sets), $\delta(a,a)=0$;  and if $a^\wedge\subset a'^\wedge$ (included sets), $\delta(a,a')=\frac{q}{q+n}$. 

This distance is very classical and is based on the Jaccard coefficient  \cite{bata:comp}. It is moreover easy though cumbersome to show that $\delta(.,.)$ is also a \emph{metric} distance.}

\begin{figure}\begin{center}\includegraphics[width=6.2cm, trim=0 20 0 0]
{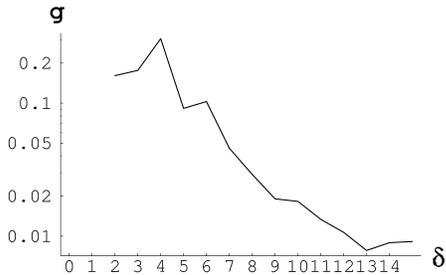}\end{center}
\caption{\label{fig:homophily} Homophilic interaction propension $\hat{g}$ with respect to $d \in\mathcal{D}=\{d_{0},...,d_{15}\}$. The y-axis is in log-scale.}
\end{figure}

\[\delta(a,a')=\frac{|(a^\wedge \setminus a'^\wedge) \cup (a'^\wedge \setminus a^\wedge)|}{|a^\wedge\cup a'^\wedge|}\]

As $\delta$ takes real values in $[0,1]$ we need to discretize $\delta$. To this end, we use a uniform partition of $[0;1[$ in $I$ intervals, to which we add the singleton $\{1\}$. We thus define a new discrete property $d$ taking values in $\mathcal{D}=\{d_{0},d_{1},...,d_{I}\}$ consisting of $I+1$ intervals: $\mathcal{D}=\left\{[0;\frac{1}{I}[;\:[\frac{1}{I};\frac{2}{I}[;...[\frac{I-1}{I};1[;\{1\}\right\}$. Finally, we obtain an empirical estimation of homophily with respect to this distance by applying Eq.~\ref{eq:g} on $d$, with $I=15$. 

The results are gathered on Fig.~\ref{fig:homophily} and show that while agents favor interactions with slightly different agents (as the initial increase suggests), they still very strongly prefer similar agents, as the clearly decreasing trend indicates (sharp decrease from $d_4$ to $d_{13}$, with $d_{4}$ being one order of magnitude larger than $d_{13}$ --- note also that $\hat{g}(d_0)=\hat{g}(d_1)=0$ because no new link appears for these distance values). 
Agents  thus display semantic homophily, a fact that fiercely advocates the necessity of taking semantic content into account when modeling such social networks.

\subsection{Correlation between degree and semantic distance}

In other words, the exponential trend of $\hat{g}$ suggests that scientists seem to choose collaborators most importantly because they are sharing interests, and less because they are attracted to well-connected colleagues, which besides actually seems to reflect agent activity. 

\begin{figure}\begin{center}\includegraphics[width=10.7cm, trim=115 00 0 0]
{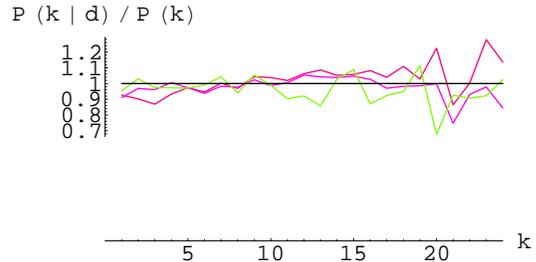}\end{center}
\caption{\label{fig:correlation} Degree and semantic distance correlation estimated through $\displaystyle\widehat{c_d}(k)=\frac{P(k|d)}{P(k)}$, plotted here for three different values of $d$: $d\in\{d_5,d_8,d_{11}\}$.}
\end{figure}

As underlined in Sec.~\ref{sec:correlation}, when building a model of such network based on degree-related and homophilic PA, one has to check whether the two properties are independent, i.e. whether or not a node of low degree is more or less likely to be at a large semantic distance of other nodes. It appears here that there is no correlation between degree and semantic distance: for a given semantic distance $d$, the probability of finding a couple of nodes including a node of degree $k$ is the same as it is for any value of $d$ --- see Fig.~\ref{fig:correlation}. To go further, we suggest 
that socio-semantic networks might be structured in communities because agents group according to similar interests, in epistemic communities \cite{roth:epis}, 
 through a mechanism involving events where agents are more or less active, and gather preferentially with respect to their interests; the former being entirely independent of the latter.

\section*{Conclusion}
Preferential attachment is the cornerstone of growth mechanisms in most recent social network formation models. This notion was established by the success of a pioneer model \cite{bara:emer} rebuilding a major stylized fact of empirical networks, the scale-free degree distribution. While PA has subsequently been widely used, few authors have tried to check or quantify the rather arbitrary assumptions on PA --- when such prospects exist, they are mostly dealing with degree-related PA or estimating PA phenomena as single parameters.   Models should go further towards empirical investigation when designing hypotheses. This would be really appealing to social scientists, who are usually not seeking 
\emph{normative} models. 
We are confident the present reluctance to measuring interaction behaviors and processes is due to the lack of a clean general framework for this purpose, which is the aim of this paper.

We thus introduced the notion of ``interaction propension'', whereby we assume that agents have an \emph{essential} preferential interaction behavior. Using this concept, we designed measurement tools for quantifying, in a dynamic network, any kind of PA with respect to any property of a single node or of a dyad of nodes --- a generalized preferential attachment. The result is a function yielding a comprehensive description of  interaction behavior related to a given property. 
In addition to clarifying PA three new features are crucial: 
 (i) properties not related to the network structure (such as homophily), (ii) correlations between properties, (iii) activity of agents and nature of interactions (e.g. modeling events, not nodes attaching to each other).
This kind of hindsight on the notion and status of PA should be useful, even for normative models.

We finally applied these tools to a particular case of socio-semantic network, a scientific collaboration network with agents linked to semantic items. While we restricted ourselves to a reduced example of two significant properties (node degree and semantic distance), measuring PA relatively to other parameters  could actually have been very relevant as well --- such as PA based on social distance for instance (shortest path length between two agents in the social network). Specifying the list of properties is nevertheless a process driven by the real-world situation  \emph{and} by the stylized facts the modeller aims at rebuilding and considers relevant for morphogenesis. 

More generally, this framework could be applied to any kind of network, as well as adapted to disconnection propensions. 
Likewise, once propensions of \emph{interaction  in the broad sense} are known, a whole class of social network morphogenesis models \cite{bogu:clas,cohe:emer,roth:bind} can be designed, with agents interacting on a growing network according to stylized interaction heuristics, heuristics precisely based on those measured empirically. \emph{In fine}, introducing more credible hypotheses based on real-case empirical measures would obviously help attract more social scientists in this promising field.

{\small \paragraph{Acknowledgements.} The author wishes to thank Cl\'emence Magnien, Matthieu Latapy and Paul Bourgine for very fruitful discussions. This work has been partially funded by the CNRS and PERSI.}

\small
\bibliography{biblio}

\begin{thebibliography}{10}

\bibitem{bara:stat}
R.~Albert and A.-L. Barab\'asi.
\newblock Statistical mechanics of complex networks.
\newblock {\em Reviews of Modern Physics}, 74:47--97, 2002.

\bibitem{bara:emer}
A.-L. Barab\'asi and R.~Albert.
\newblock Emergence of scaling in random networks.
\newblock {\em Science}, 286:509--512, 1999.

\bibitem{bara:evol}
A.-L. Barab\'asi, H.~Jeong, R.~Ravasz, Z.~Neda, T.~Vicsek, and T.~Schubert.
\newblock Evolution of the social network of scientific collaborations.
\newblock {\em Physica A}, 311:590--614, 2002.

\bibitem{bata:comp}
V.~Batagelj and M.~Bren.
\newblock Comparing resemblance measures.
\newblock {\em Journal of Classification}, 12(1):73--90, 1995.

\bibitem{bogu:clas}
M.~Boguna and R.~Pastor-Satorras.
\newblock Class of correlated random networks with hidden variables.
\newblock {\em Physical Review E}, 68:036112, 2003.

\bibitem{bogu:mode}
M.~Boguna, R.~Pastor-Satorras, A.~Diaz-Guilera, and A.~Arenas.
\newblock Models of social networks based on social distance attachment.
\newblock {\em Physical Review E}, 70:056122, 2004.

\bibitem{cald:scal}
G.~Caldarelli, A.~Capocci, P.~D.~L. Rios, and M.~A. Munoz.
\newblock Scale-free networks from varying vertex intrinsic fitness.
\newblock {\em Physical Review Letters}, 89(25):258702, 2002.

\bibitem{cata:asso}
M.~Catanzaro, G.~Caldarelli, and L.~Pietronero.
\newblock Assortative model for social networks.
\newblock {\em Physical Review E}, 70:037101, 2004.

\bibitem{cohe:emer}
P.~Cohendet, A.~Kirman, and J.-B. Zimmermann.
\newblock Emergence, formation et dynamique des r\'eseaux -- mod\`eles de la
  morphogen\`ese.
\newblock {\em Revue d'Economie Industrielle}, 103(2-3):15--42, 2003.

\bibitem{colizza:network}
V.~Colizza, J.~R. Banavar, A.~Maritan, and A.~Rinaldo.
\newblock Network structures from selection principles.
\newblock {\em Physical Review Letters}, 92(19):198701, 2004.

\bibitem{eise:pref}
E.~Eisenberg and E.~Y. Levanon.
\newblock Preferential attachment in the protein network evolution.
\newblock {\em Physical Review Letters}, 91(13):138701, 2003.

\bibitem{erdo:rand}
P.~Erd\"os and A.~R\'enyi.
\newblock On random graphs.
\newblock {\em Publicationes Mathematicae}, 6:290--297, 1959.

\bibitem{fabr:heur}
A.~Fabrikant, E.~Koutsoupias, and C.~H. Papadimitriou.
\newblock Heuristically optimized trade-offs: A new paradigm for power laws in
  the internet.
\newblock In {\em ICALP '02: Proceedings of the 29th International Colloquium
  on Automata, Languages and Programming}, pages 110--122, London, UK, 2002.
  Springer-Verlag.

\bibitem{guil:bipa}
J.-L. Guillaume and M.~Latapy.
\newblock Bipartite structure of all complex networks.
\newblock {\em Information Processing Letters}, 90(5):215--221, 2004.

\bibitem{jeong:measurin}
H.~Jeong, Z.~N\'eda, and A.-L. Barabasi.
\newblock Measuring preferential attachment for evolving networks.
\newblock {\em Europhysics Letters}, 61(4):567--572, 2003.

\bibitem{jin:stru}
E.~M. Jin, M.~Girvan, and M.~E.~J. Newman.
\newblock The structure of growing social networks.
\newblock {\em Physical Review E}, 64(4):046132, 2001.

\bibitem{lata:bas2}
M.~Latapy, C.~Magnien, M.~Mariadassou, and C.~Roth.
\newblock A basic toolbox for the analysis of dynamics of growing networks.
\newblock Forthcoming.

\bibitem{mcph:homo}
M.~McPherson and L.~Smith-Lovin.
\newblock Birds of a feather: Homophily in social networks.
\newblock {\em Annual Review of Sociology}, 27:415--440, 2001.

\bibitem{newm:clus}
M.~E.~J. Newman.
\newblock Clustering and preferential attachment in growing networks.
\newblock {\em Physical Review Letters E}, 64(025102), 2001.

\bibitem{newm:stru}
M.~E.~J. Newman.
\newblock The structure of scientific collaboration networks.
\newblock {\em PNAS}, 98(2):404--409, 2001.

\bibitem{powe:grow}
W.~W. Powell, D.~R. White, K.~W. Koput, and J.~Owen-Smith.
\newblock Network dynamics and field evolution: The growth of
  interorganizational collaboration in the life sciences.
\newblock {\em American Journal of Sociology}, 110(4):1132--1205, 2005.

\bibitem{rama:self}
J.~J. Ramasco, S.~N. Dorogovtsev, and R.~Pastor-Satorras.
\newblock Self-organization of collaboration networks.
\newblock {\em Physical Review E}, 70:036106, 2004.

\bibitem{redn:cita}
S.~Redner.
\newblock Citation statistics from 110 years of physical review.
\newblock {\em Physics Today}, 58:49--54, 2005.

\bibitem{roth:bind}
C.~Roth and P.~Bourgine.
\newblock Binding social and cultural networks: a model.
\newblock arXiv.org e-print archive, nlin.AO/0309035, 2003.

\bibitem{roth:epis}
C.~Roth and P.~Bourgine.
\newblock Epistemic communities: Description and hierarchic categorization.
\newblock {\em Mathematical Population Studies}, 12(2):107--130, 2005.

\bibitem{BrianSkyrms08012000}
B.~Skyrms and R.~Pemantle.
\newblock {A dynamic model of social network formation}.
\newblock {\em PNAS}, 97(16):9340--9346, 2000.

\bibitem{snij:stat}
T.~A. Snijders.
\newblock The statistical evaluation of social networks dynamics.
\newblock {\em Sociological Methodology}, 31:361--395, 2001.

\bibitem{sode:gene}
B.~S\"oderberg.
\newblock A general formalism for inhomogeneous random graphs.
\newblock {\em Physical Review E}, 68:026107, 2003.

\bibitem{stef:pref}
H.~Stefancic and V.~Zlatic.
\newblock Preferential attachment with information filtering--node degree
  probability distribution properties.
\newblock {\em Physica A}, 350(2-4):657--670, 2005.

\end{thebibliography}
\bibliographystyle{abbrv}

\end{document}